\documentclass[preprint,preprintnumbers,superscriptaddress,amsmath,amssymb]{revtex4}

\usepackage{graphicx}
\usepackage{dcolumn}
\usepackage{bm}
\usepackage{amsmath}
\newcommand{\be}{\begin{equation}}
\newcommand{\ee}{\end{equation}}
\newcommand{\ba}{\begin{eqnarray}}
\newcommand{\ea}{\end{eqnarray}}

\begin{document}

\preprint{APS preprint}

\title{Anomalous Power Law Distribution \\of Total Lifetimes of
Aftershocks Sequences}

\author{A. Saichev}
\affiliation{Mathematical Department,
Nizhny Novgorod State University, Gagarin prosp. 23,
Nizhny Novgorod, 603950, Russia}
\affiliation{Institute of Geophysics and Planetary Physics,
University of California, Los Angeles, CA 90095}

\author{D. Sornette}
\affiliation{Institute of Geophysics and Planetary Physics,
University of California, Los Angeles, CA 90095}
\affiliation{Department of Earth and Space Sciences, University of
California, Los Angeles, CA 90095\label{ess}}
\affiliation{Laboratoire de Physique de la Mati\`ere Condens\'ee,
CNRS UMR 6622 and Universit\'e de Nice-Sophia Antipolis, 06108
Nice Cedex 2, France}
\email{sornette@moho.ess.ucla.edu}

\date{\today}

\begin{abstract}
We consider a general stochastic branching process, which
is relevant to earthquakes, and study the distributions of
global lifetimes of the branching processes. In the
earthquake context, this amounts to the distribution of
the total durations of aftershock sequences including
aftershocks of arbitrary generation numbers. Our results
extend previous results on the distribution of the total
number of offsprings (direct and indirect aftershocks in
seismicity) and of the total number of generations before
extinction. We consider a branching model of triggered
seismicity, the ETAS (epidemic-type aftershock sequence)
model which assumes that each earthquake can trigger other
earthquakes (``aftershocks''). An aftershock sequence
results in this model from the cascade of aftershocks of
each past earthquake. Due to the large fluctuations of the
number of aftershocks triggered directly by any earthquake
(``productivity'' or ``fertility''), there is a large variability of the total
number of aftershocks from one sequence to another, for
the same mainshock magnitude. We study the regime where
the distribution of fertilities $\mu$ is characterized by
a power law $\sim 1/\mu^{1+\gamma}$ and the bare Omori law
for the memory of previous triggering mothers decays
slowly as  $\sim 1/t^{1+\theta}$, with $0 < \theta <1$
relevant for earthquakes. Using the tool of generating
probability functions and a quasistatic approximation
which is shown to be exact asymptotically for large
durations, we show that the density distribution of total
aftershock lifetimes scales as $\sim
1/t^{1+\theta/\gamma}$ when the average branching ratio is critical ($n=1$).
The coefficient $1<\gamma = b/\alpha<2$
quantifies the interplay between the exponent $b \approx
1$ of the Gutenberg-Richter magnitude distribution $ \sim
10^{-bm}$ and the increase $\sim 10^{\alpha m}$ of the
number of aftershocks with the mainshock magnitude $m$ (productivity)
with $\alpha \approx 0.8$. The renormalization of the bare
Omori decay law $\sim 1/t^{1+\theta}$ into $\sim
1/t^{1+\theta/\gamma}$ stems from the nonlinear
amplification due to the heavy-tailed distribution of
fertilities and the critical nature of the branching cascade process.
In the subcritical case $n<1$, the cross-over for $\sim
1/t^{1+\theta/\gamma}$ at early times to $\sim 1/t^{1+\theta}$
at longer times is described. More generally, our results apply to any
stochastic branching process with a power-law distribution
of offsprings per mother and a long memory.
\end{abstract}


\maketitle

\section{Introduction}

We study the distribution of the total duration of an
aftershock sequence, for a class of branching processes
\cite{Athreya,Sankaranarayanan} appropriate in particular
for modeling earthquake aftershock sequences. The
noteworthy particularity and challenging property of this
class of branching processes is that the variance of the
number of progenies in direct lineage from the mother is
mathematically infinite. In addition, a long-time (power
law) memory of the impact of a mother on triggering her
first-generation daughters gives rise to subdiffusion
\cite{etasdif,HOS03} and non-mean field behavior in the
distributions of the total number of aftershocks per
mainshock and of the total number of generations before
extinctions \cite{Saichevetal04}. Here, we add on these
previous works but showing that the distribution of the
total duration of an aftershock sequence is extremely
long-tailed: the very heavy-tailed nature of the
distribution of the durations of aftershock sequences
predicted by this simple model may explain the large
variability of the lifetimes of observed aftershock
sequences and is compatible with the observation that felt
aftershocks of the great Mino-Owari (1891) Japanese
earthquake, that inspired Omori's statistical rate model,
have persisted at a rate consistent with the Omori law for
100 years \cite{Utsuetal95}.

Our results may also be of interest to other systems
which are characterized by branching processes with
a broad power-law distribution of fertilities, such as
epidemic transmission of diseases, and more generally transmission
processes involving avalanches spreading on networks such as the
World Wide Web, cellular metabolic network, ecological food webs,
social networks and so on, as a consequence of the well-documented power law
distribution of connectivities among nodes. Our results are thus relevant
to systems in which the number of offsprings may be large due to
long-range interactions, long-memory effects or large deviation processes.

\section{The Epidemic-Type Aftershock Sequence (ETAS) branching model
of earthquakes
with long memory \label{laws}}

We consider a general branching process in which each progenitor or
mother (mainshock)
is characterized by its conditional average number
\be
N_m \equiv \kappa \mu(m)
\label{avenun}
\ee
of children (triggered
events or aftershocks of first generation), where
\be
\mu(m) = 10^{\alpha (m-m_0)}~,
\label{mudef}
\ee
is a mark associated with an
earthquake of magnitude $m \geq m_0$ (in the language of
``marked point processes''),  $\kappa$ is a constant factor
and $m_0$ is the minimum magnitude of earthquakes capable of triggering
other earthquakes. The meaning of the term ``conditional average''
for $N_m$ is the following: for
a given earthquake of magnitude $m$ and therefore
of mark $\mu(m)$, the number $r$ of its daughters of first generation
are drawn at random according to the Poissonian statistics
\be
p_\mu(r)= \frac{N_m^r}{r!}\,e^{-N_m} =
\frac{(\kappa\mu)^r}{r!}\,e^{-\kappa\mu}\,.
\label{3}
\ee
Thus, $N_m$ is the expectation of the number of daughters of first
generation, conditioned on a fixed magnitude $m$ and mark $\mu(m)$.
The expression (\ref{mudef}) for $\mu(m)$ is chosen
in such a way that it reproduces the empirical dependence of the
average number of aftershocks triggered directly by an earthquake
of magnitude $m$ (see \cite{alpha} and references therein).
Expression (\ref{avenun}) with (\ref{mudef}) gives the so-called
productivity law of a given mother as a function of its magnitude.

In addition, we use the well-known Gutenberg-Richter (GR)
density distribution of earthquake magnitudes
\be
p(m) = b ~\ln (10)~ 10^{-b (m-m_0)}~,~~~~m \geq m_0~, \label{GR}
\ee
such that $\int_m^{\infty} p(x) dx$
gives the probability that an earthquake has a
magnitude equal to or larger than $m$. This magnitude
distribution $p(m)$ is assumed to be independent on the
magnitude of the triggering earthquake, i.e., a large
earthquake can be triggered by a smaller one
\cite{alpha,Forexp}.

Combining (\ref{GR}) and (\ref{mudef}), we see that the
earthquake marks $\mu$ and therefore the conditional average
number $N_m$ of daughters of first generation are
distributed according to a power law
\be
p_{\mu}(\mu) = {\gamma \over \mu^{1+\gamma}}~,
~~~1 \leq \mu < +\infty, ~~~~~\gamma = b/\alpha~.
\label{aera}
\ee
Note that $p_{\mu}(\mu)$ is normalized: $\int_1^{+\infty} d\mu
~p_{\mu}(\mu)=1$.
For earthquakes, $b \approx 1$ almost universally and $\alpha \approx
0.8$ \cite{alpha},
giving $\gamma \approx 1.25$. The fact that $1 < \gamma <2$ implies that the
mathematical expectation of $\mu$ and therefore of $N_m$
(performed over all possible magnitudes) is finite but its variance
is infinite.

For a fixed $\gamma$, the coefficient $\kappa$ then controls the value of the
average number $n$ of children of first generation per mother:
\be
n = \langle N_m \rangle = \kappa \langle \mu \rangle = \kappa {\gamma
\over \gamma -1}~,
\label{mgmlele}
\ee
where the average $\langle N_m \rangle$ is taken over all mothers' magnitudes
drawn from the GR law.
In the terminology of branching processes, $n$ is called the branching ratio.
For $n<1$, there are on average less than one child per mother:
this corresponds to transient (sub-critical) branching processes
with finite lifetimes with probability one. For $n>1$,
there are more than one child per mother: this corresponds
to explosive (super-critical) branching processes with a number of events
growing exponentially with time. The value $n=1$ of exactly one child
per mother
on average is the critical point separating the two regimes.

Finally, we assume that a given event (the ``mother'') of
magnitude $m\geq m_0$ occurring at time $t_i$ gives birth
to other events (``daughters'') of first generation in the
time interval between $t$ and $t+dt$  at the rate
\be
\phi_\mu(t) = N_{m}~\Phi(t-t_i) = N_{m} ~
{\theta~c^{\theta} \over (t+c)^{1+\theta}}~H(t)
\label{eq1dffrf}
\ee
where $0<\theta<1$, $H(t)$ is the
Heaviside function, $c$ is a regularizing time scale that
ensures that the seismicity rate remains finite close to
the mainshock and $N_m$ is given by (\ref{avenun}). The
time decay rate (\ref{eq1dffrf}) is called the ``direct
Omori law''  \cite{SS99,HS02}. Due to the process of
cascades of triggering by which a mother triggers
daughters which then trigger their own daughters and so
on, the direct Omori law (\ref{eq1dffrf}) is renormalized
into a ``dressed'' or ``renormalized'' Omori law
\cite{SS99,HS02}, which is the one observed empirically.

Expressions (\ref{avenun},\ref{mudef},\ref{GR},\ref{eq1dffrf}) define the
Epidemic-Type Aftershock Sequence model of triggered seismicity
introduced by Ogata in the present form
\cite{Ogata} and  by Kagan and Knopoff in a slightly different form
\cite{KK81}.

\section{General formalism in term of generating functions}

Since we are interested in characterizing the distribution of the
random times at which an aftershock sequence triggered by a given
mainshock terminates, we take the time of the mainshock
of magnitude $m$ at the origin $t=0$ and we do not consider the
effect of earlier earthquakes. This is warranted by
the fact that sequences of earthquakes generated by different mainshocks
are independent in the ETAS branching model.

\subsection{First generation aftershocks}

Let us first discuss some more detailed statistical
description of first generation aftershocks. Each
aftershock arising independently from another preceding
aftershock itself born at the random time $t_i$ has its
birth time possessing the probability density function
(PDF) $\Phi(t-t_i)$ defined in (\ref{eq1dffrf}) and
cumulative distribution function (CDF)
$b(t)=\int_0^t\Phi(t')\,dt'$. Here and everywhere below,
the dimensionless time $t/c$ is used and we replace $t$ by
$t/c$, with the understanding that $t$ or $\tau$ means
$t/c$ when needed. It is convenient to introduce the
complementary CDF of first generation aftershocks \be
a(t)=1-b(t)=\frac{1}{(t+1)^\theta}\,. \label{abequiv} \ee

Let us consider a mainshock with mark $\mu$ that triggers
exactly $r$ aftershocks of first-generation arising at the
moments $(t_1,t_2,\dots,t_r)$. Then, the CDF of the time
$T(\mu|r)$ of the last arising aftershock is equal to \be
\text{P}_\mu(t|r)= \text{Pr\,}\left(T(\mu|r)=
\max\{t_1,t_2,\dots,t_r\}<t\right)= [b(t)]^r\,. \ee
Averaging this CDF over the random first-generation
aftershock numbers $r$ at fixed $\mu$ weighted by their
probability $p_\mu(r)$ given by (\ref{3}) yields the CDF
$\text{P}_\mu(t)$ for the total duration $T(\mu)$ of the
first-generation aftershocks \be \text{P}_\mu(t)=
\text{Pr\,}(T(\mu)<t)= G_\mu[b(t)]\,. \label{4} \ee Here,
$G_\mu(z)= \sum_{r=0}^\infty p_\mu(r)z^r$ is the
generating probability function (GPF) of the number of
first-generation aftershocks. For the Poissonian
statistics (\ref{3}), it is equal to \be G_\mu(z)=
e^{\kappa\mu(z-1)}\,. \label{5}
\ee
This leads to the well-known relation
\be
\text{P}_\mu(t)=e^{-\kappa\mu\,a(t)}\,. \ee

In the ETAS model, the Gutenberg-Richter distribution (\ref{GR})
of magnitudes together with the productivity law (\ref{mudef})
implies the power law (\ref{aera}) for the marks $\mu$.
Averaging over all possible mainshock magnitude thus amounts
to averaging (\ref{4}) over all possible $\mu$'s.
The CDF of durations $T$ of first-generation aftershocks
generated by some mother of arbitrary magnitude arising at time $t=0$
is equal to
\be
\text{P\,}(t)=G[b(t)]\,,
\label{7}
\ee
where $G(z)=\langle G_\mu(z)\rangle$ is the average of $G_\mu[b(t)]$ over
the random magnitudes $m$ (or equivalently random marks $\mu$)
In the relevant case of the Poissonian GPF (\ref{5}) and using (\ref{aera}),
we obtain
\be
G(z)=\gamma \kappa^\gamma (1-z)^\gamma\,
\Gamma(-\gamma,\kappa(1-z))\,,
\label{8}
\ee
where $\Gamma(x,y)$ is the incomplete Gamma function and
$\gamma=b/\alpha$. For real aftershocks,
$1<\gamma<2$ and a typical value is $\gamma \approx 1.25$. Then, it
is easy to show that the first terms of $G(z)$ in a power
expansion with respect to $1-z$ are
\be
G(z)\simeq 1-n(1-z)+\beta(1-z)^\gamma\,, \qquad
1<\gamma<2 \,,
\label{9}
\ee
with $n$ given by (\ref{mgmlele}) and
\be
\beta=n^\gamma \left(\frac{\gamma-1}{\gamma}\right)^\gamma
\frac{\Gamma(2-\gamma)}{\gamma-1}\,.
\label{betadef}
\ee

\subsection{All generation aftershocks}

In the ETAS model, any event (the initial mother or any aftershock,
whatever its generation number)
triggers its aftershocks of first-generation
in a statistically independent and equivalent manner,
according to the laws given in section \ref{laws}.
This gives the possibility of obtaining
closed equations for the CDF of the total
duration of aftershocks triggering processes.

Let ${\cal T}$ be the random waiting time between a mainshock and
one of his first-generation aftershocks, chosen arbitrarily.
The PDF of ${\cal T}$ is nothing but $\Phi(t)$ defined in
(\ref{eq1dffrf}).
Let $\mathbb{T}$ be the random duration of the aftershock branching process
triggered by this first-generation aftershock. The CDF of
$\mathbb{T}$ is denoted
$\mathbb{P}(t)$. Then, the total
duration, measured since the mainshock,
of the sequence of aftershocks generated by this pointed out
first-generation aftershock is ${\cal T}+\mathbb{T}$. The CDF
$\mathbb{F}(t)$ of this sum
is therefore the convolution
\be
\mathbb{F}(t)=\Phi(t)\otimes\mathbb{P}(t)\,.
\label{10}
\ee
Replacing in (\ref{4}) $b(t)$ by $\mathbb{F}(t)$
and taking into account the equality (\ref{5}), we obtain
the CDF of the total duration $\mathbb{T}(\mu)$ of
a sequence of aftershocks over all generations of a given event
of mark $\mu$ that occurred at $t=0$:
\be
\mathbb{P}_\mu(t)= \text{Pr\,}(\mathbb{T}(\mu)<t)=
e^{-\kappa \mu \mathbb{R}(t)}~,
\label{11}
\ee
where
\be
\mathbb{R}(t)=1-\mathbb{F}(t)
\label{RF}
\ee
is the complementary to the $\mathbb{F}(t)$ CDF defined in (\ref{10}).
Correspondingly, replacing in (\ref{7}) $\text{P\,}(t)$ by
$\mathbb{P}(t)$ and $b(t)$ by $\mathbb{F}(t)$, we obtain
the self-consistent equation for the CDF $\mathbb{F}(t)$
\be
\mathbb{P}(t)=G[\mathbb{F}(t)] = G\left[\Phi(t)\otimes\mathbb{P}(t)\right]\,.
\label{12}
\ee

It is convenient to rewrite (\ref{12}) as
\be
\mathbb{R}(t)-\mathbb{Q}(t)=
\Omega\left[\mathbb{R}(t)\right]\,,
\label{13}
\ee
where $\mathbb{Q}(t)=1-\mathbb{P}(t)$ and
\be
\Omega(z)=G(1-z)+z-1\,.
\label{14}
\ee
For our subsequent analysis, expression (\ref{13}) is more convenient than
equation (\ref{12}) for the following reasons. First of all, instead
of the CDF's $\mathbb{P}(t)$ and $\mathbb{F}(t)$
entering in (\ref{12}), equation (\ref{13})
is expressed in terms of the complementary CDF's $\mathbb{Q}(t)$ and
$\mathbb{R}(t)$ which both tend to zero for $t\to\infty$. In addition,
the function $\Omega(z)$ also tends to zero for $z\to
0$. This gives the possibility of extracting the influence of
the nonlinear terms of the GPF $G(z)$ on the asymptotic behavior
of the solution for $t\to\infty$. Indeed, at
least for $\gamma \lesssim 1.5$, the GPF $G(z)$ is very
precisely described by the truncated series (\ref{9}).
The corresponding series for $\Omega(z)$ is
\be
\Omega(z)\simeq (1-n)z+\beta z^\gamma ~,
\label{15}
\ee
which reduces to a pure power law in the critical case $n=1$:
\be
\Omega(z)\simeq \beta z^\gamma\,.
\label{16}
\ee
Correspondingly, in the critical case $n=1$ and most
important for earthquake applications for which $1<\gamma<2$ holds,
equation (\ref{13}) has the form
\be
\mathbb{R}(t)-\mathbb{Q}(t)=\beta \mathbb{R}^\gamma(t)\,.
\label{13bis}
\ee
The exact auxiliary function $\Omega(z)$
defined by (\ref{14}) for $n=1$
and its power approximation (\ref{16}) for $\gamma=1.25$
are shown in Fig.~1.

Our goal is now to solve (\ref{13}) and in particular (\ref{13bis})
to explore in details the statistical properties of the durations of
aftershocks sequences, resulting from cascades of triggered events.

\section{Fractional order differential equation for the
complementary CDF $\mathbb{R}(t)$}

In order to exploit equation (\ref{13}), we first need to
express $\mathbb{Q}(t)$ as a function of $\mathbb{R}(t)$.
For this, we note that expression (\ref{10}) is equivalent
to
\be
\mathbb{R}(t)=a(t)+\Phi(t)\otimes \mathbb{Q}(t)\,,
\ee
as can be seen from direct substitutions using
(\ref{abequiv}), (\ref{RF}) and
$\mathbb{Q}(t)=1-\mathbb{P}(t)$. Applying the Laplace
transform to both sides of this equality, one gets \be
\hat{\mathbb{Q}}(s)=
\frac{\hat{\mathbb{R}}(s)}{\hat{\Phi}(s)}-
\frac{1-\hat{\Phi}(s)}{s\,\hat{\Phi}(s)}\,, \label{17} \ee
where \be \hat{\Phi}(s)=\int_0^\infty \Phi(t) e^{-st}\,
dt= \theta\, (cs)^\theta\, e^{cs}\, \Gamma(-\theta,s)\,,
\ee where we have made the correspondence $t \to t/c$
explicit (where $c$ is defined in (\ref{eq1dffrf})).
We shall be interested in the probability
distribution of the durations of total sequences of
aftershocks for durations much larger than $c$. In this
case, one can replace $\hat{\Phi}(s)$ by its asymptotics
for small $s$
\be
\hat{\Phi}(s)\simeq 1-\delta
(c\,s)^\theta \simeq \frac{1}{1+\delta
(c\,s)^\theta}\,,\qquad c\,s\ll 1\,,
\ee
where
$\delta=\Gamma(1-\theta)$.  Substituting it into
(\ref{17}) leads to
\be
\hat{\mathbb{Q}}(s)=
\left[1+\delta(c\,s)^\theta\right]
\hat{\mathbb{R}}(s)-\delta\, c^\theta s^{\theta-1}\,,
\ee
which is equivalent, under the inverse Laplace transform,
to the fractional order differential equation
\be
\mathbb{Q}(t)=\mathbb{R}(t)+\delta\,c^\theta\,
\frac{d^\theta \mathbb{R}(t)}{dt^\theta}-
\left(\frac{c}{t}\right)^\theta\,.
\ee
Equation (\ref{13})
thus yields the following fractional order differential
equation for $\mathbb{R}(t)$ (going back to the reduced
time variable $\tau=t/c$) \be \delta \frac{d^\theta
\mathbb{R}}{d\tau^\theta}+\Omega(\mathbb{R})=
\tau^{-\theta}\,. \label{18} \ee In particular in the
critical case $n=1$, using the power approximation
(\ref{16}), we obtain \be \delta\, \frac{d^\theta
\mathbb{R}}{d\tau^\theta}+\beta\,
\mathbb{R}^\gamma=\tau^{-\theta}\,. \label{19} \ee

Note that the nonlinear fractional order
differential equation (\ref{18}) is exact for $\Phi(t)$ given by
\be
\Phi(t)=\frac{1}{\delta^{1/\theta}} \Phi_\theta\left(
\frac{t}{\delta^{1/\theta}}\right)\,,
\label{20}
\ee
where $\Phi_\theta(t)$ is the fractional exponential
distribution possessing the following Laplace transform
\be
\hat{\Phi}_\theta(s)=\frac{1}{1+s^\theta}\,,
\ee
which has the integral representation
\be
\Phi_\theta(\tau)=\int_0^\infty \frac{1}{x}
\exp\left(-\frac{\tau}{x}\right)\,\xi_\theta(x)\, dx\,,
\label{21}
\ee
where
\be
\xi_\theta(x)=\frac{1}{\pi x}\,
\frac{\sin(\pi\theta)}{x^\theta+x^{-\theta}+2
\cos(\pi\theta)}\,.
\label{22}
\ee
One can interpret (\ref{21}) as the decomposition of the fractional
exponential law into regular exponential distributions,
and $\xi_\theta(x)$ given by (\ref{22}) as the ``spectrum'' of their mean
characteristic decay time $x$. For $\theta \to 1$, the spectrum
(\ref{22}) weakly
converges to the delta-function $\delta(x-1)$ and the fractional
exponential law transforms into the regular exponential
distribution $\Phi_1(\tau)=e^{-\tau}$.
For $\theta=1/2$, there is an explicit expression for
the fractional exponential distribution
\be
\Phi_{1/2}(\tau)=\sqrt{\frac{1}{\pi \tau}}-e^\tau\,
\text{erfc\,}(\sqrt{\tau})\,.
\ee
It is easy to show that the asymptotics of the fractional
exponential distribution are
\be
\Phi_\theta(\tau)\simeq
\frac{\tau^{\theta-1}}{\Gamma(\theta)} \qquad (\tau\ll 1
)\,, \qquad \Phi_\theta(\tau)\simeq \frac{\theta\,
\tau^{-\theta-1}}{\Gamma(1-\theta)}\qquad (\tau\gg 1)\,.
\ee

Fig.~2 shows a log-log plot of the Omori law $\Phi(t)$ defined
in (\ref{eq1dffrf}) and of the
corresponding fractional exponential distribution (\ref{20}) as
a function of the reduced time $\tau=t/c$
and for $\theta=1/2$, demonstrating the closeness of
these two distributions.

\section{Exactly soluble case: pure exponential Omori law}

Before addressing the case of interest of earthquakes where
the direct Omori law $\Phi(t)$ is a power law with exponent $0 < \theta < 1$,
it is instructive to present the solution for the case where $\Phi(t)$ is an
exponential. In this case, an exact solution can be obtained in close form.
This exact solution will be useful to check the
quasistatic and dynamical linearization approximations
developed below to solve the difficult case where
$\Phi(t)$ is a power law with exponent $0 < \theta < 1$.

We write the exponential direct Omori law in non-reduced time as
\be
\Phi(t)= \frac{1}{c} \exp\left(-\frac{t}{c}\right)
\qquad \Rightarrow \qquad \hat{\Phi}(s)=\frac{1}{1+c
s}\,,
\label{23}
\ee
so that equation (\ref{17}) transforms into
\be
\hat{\mathbb{Q}}(s)=(1+cs)\hat{\mathbb{R}}(s)-c\,.
\ee
After inverse Laplace transform, we get
\be
\mathbb{Q}(t)=\mathbb{R}(t)+c\frac{d
\mathbb{R}(t)}{dt}-c\delta(t)\,,
\ee
and equation (\ref{13}) takes the form
\be
c\frac{d \mathbb{R}(t)}{dt}+
\Omega\left[\mathbb{R}(t)\right]= c\delta(t)\,,
\ee
or, in the more traditional form of a Cauchy problem
\be
\frac{d \mathbb{R}}{d\tau}+
\Omega\left[\mathbb{R}\right]=0\,, \qquad
\mathbb{R}(\tau=0)=1\,.
\label{24}
\ee
The numerical solution of (\ref{24}) is easy to obtain. In addition,
using for $\Omega(z)$ the series approximation (\ref{15}), one obtains
the analytical solution of the Cauchy problem (\ref{24}) under the form
\be
\mathbb{R}=\left[\left(1+\frac{\beta}{1-n}\right)
\exp\left((1-n)\frac{\tau}{g}\right)-
\frac{\beta}{1-n}\right]^{-g}\,,
\label{25}
\ee
where $g=1/(\gamma-1)$. In particular, in the critical case
$n=1$, this leads to
\be
\mathbb{R}=\left(1+\frac{\beta}{g}\,\tau\right)^{-g}\,.
\label{26}
\ee
Fig.~3 shows the numerical solution of
equation (\ref{24}) together with its analytical solution (\ref{25}) obtained
using the polynomial approximation (\ref{15}) of the function
$\Omega(z)$ defined in (\ref{14}), for $\gamma=1.25$ and $n=0.99$. It is
seen that curves are very close each other.

Note that, in the subcritical case $n<1$, there is a crossover from the
power law (\ref{26}) at early times
which is characteristic of the critical regime $n=1$, to
an exponential decay at long times of the complementary CDF $\mathbb{R}$.

\section{Dynamical linearization and quasistatic approximations to
obtain the asymptotic tail of the distribution of total aftershock durations}

\subsection{Linear approximation}

To obtain some rough estimate of the complementary CDF
$\mathbb{R}(t)$, let us consider the linearized version of the
fractional order differential equation (\ref{18})
\be
\delta \frac{d^\theta
\mathbb{R}}{d\tau^\theta}+\eta\,\mathbb{R}= \tau^{-\theta} ~,
\label{27}
\ee
where the following linearization has been used
\be
\Omega[\mathbb{R}]\simeq \eta\,\mathbb{R}\,, \qquad
\eta=\Omega(1)=G(0)\,.
\label{28}
\ee
The Laplace transform of the solution of the linearized equation (\ref{27})
has the form
\be
\hat{\mathbb{R}}(s)=\frac{\delta
s^{\theta-1}}{\eta+\delta s^\theta}\,.
\ee
The corresponding complementary CDF is equal to
\be
\mathbb{R}=E_\theta\left(-\frac{\eta}{\delta}\,
\tau^{\theta}\right)\,,\qquad \delta=\Gamma(1-\theta)\,,
\label{29}
\ee
where $E_\theta(z)$ is the Mittag-Leffler function. Its
integral representation is
\be
E_\theta(-x)=\frac{x}{\pi}\sin\pi\theta\int_0^\infty
\frac{y^{\theta-1} e^y\,dy}{y^{2\theta}+x^2+ 2 x
y^\theta \cos\pi\theta}\qquad (x>0)\,.
\label{30}
\ee
In particular for $\theta=1/2$, it is equal to
\be
E_{1/2}(-x)=e^{x^2}\text{erfc\,}(x)\,.
\label{31}
\ee
Its asymptotics reads
\be
E_\theta(-x)\sim \frac{1}{x\delta} \qquad (x\to\infty)~,
\label{32}
\ee
which is already very precise for
$x\gtrsim 2$.

The suggested dynamical linearization approach consists in
replacing the factor $\eta$ in (\ref{28}) by
\be
\eta(\mathbb{R})=\frac{\Omega(\mathbb{R})}{\mathbb{R}}\,.
\label{39}
\ee
to correct for the nonlinear decay of the relaxation of
the complementary CDF $\mathbb{R}$ as a function of time.
It is interesting to check the validity of this dynamical
linearization procedure for the exactly solvable
exponential Omori law (\ref{23}). In this case, the solution of
the linearized equation (\ref{24}) is
\be
\mathbb{R}=e^{-\eta\tau}\,.
\ee
Substituting here (\ref{39}) for $\eta$, we obtain in the critical case
the following
transcendent equation
\be
\mathbb{R}=\exp(-\tau\,\beta\, \mathbb{R}^{\gamma-1})\,.
\ee
Its solution is equal to
\be
\mathbb{R}=\left(\frac{Y(x)}{x}\right)^g\,,
\label{43}
\ee
where
\be
g=\frac{1}{\gamma-1}\,, \qquad x=\frac{\tau\beta}{g}\,,
\ee
and $Y(x)$ is the solution of the transcendent equation $Y\,e^Y=x$.
For $x>2$, there is very precise approximate solution of
this equation:
\be
Y(x)\simeq \ln x\left[1+(1+\ln x)\left(1- \sqrt{1+
\frac{2\ln(\ln x)}{(1+\ln x)^2}}\right)\right]\sim \ln x\,.
\ee
Thus, for large $x$, the main asymptotics of the dynamical
linearization approximation (\ref{43}) of the Cauchy problem (\ref{24})
differs from the main asymptotics $\mathbb{R}\sim
x^{-g}$ of the exact solution (\ref{26})
only by logarithmic correction $\ln^g x$.

\subsection{Quasistatic approximation}

Close inspection of the complementary CDF (\ref{29}) and its
asymptotics
\be
\mathbb{R}\simeq \frac{1}{\eta\tau^\theta}\,, \qquad
\tau\gtrsim \tau^*\,, \qquad \tau^*=
\left(\frac{2\delta}{\eta}\right)^{1/\theta}
\label{33}
\ee
derived from relation (\ref{32}) gives us a hint of how
to approach the solution
of the nonlinear fractional order differential equation
(\ref{18}) and (\ref{19}) by using a quasistatic approximation. Indeed,
notice that the asymptotics (\ref{33}) is solution of the truncated
equation (\ref{27})
\be
\eta\,\mathbb{R}=\tau^{-\theta} ~,
\label{34}
\ee
where we omitted the fractional order derivative term.

Applying this same quasistatic approximation to
the nonlinear fractional order differential equation (\ref{18}) gives
the approximate equality
\be
\Omega[\mathbb{R}]\simeq \tau^{-\theta}\,.
\label{35}
\ee
In particular, in the critical case $n=1$ for which
$\Omega(z)\simeq \beta z^\gamma$, we have
$\beta\,\mathbb{R}^\gamma\simeq\tau^{-\theta}$,
or equivalently
\be
\mathbb{R}\simeq \beta^{-1/\gamma}\,
\tau^{-\theta/\gamma}\,.
\label{36}
\ee
Expression (\ref{36}) will lead to your main result (\ref{49}) below.

The validity of this quasistatic approximation is checked
by calculating the derivation of fractional order $\theta$
of the approximate solution (\ref{36}). Using the standard
tabulated formula of fractional order analysis \be
\frac{d^\theta \tau^p}{d\tau^\theta}=
\frac{\Gamma(1+p)}{\Gamma(1+p-\theta)}\,
\tau^{p-\theta}\,, \label{37} \ee we obtain \be \delta
\frac{d^\theta \mathbb{R}}{d\tau^\theta}\simeq
-\beta^{-1/\gamma}\, \frac{\theta}{\gamma+1}\,
\text{B}\left(-\theta,-\frac{\theta}{\gamma}\right)\,
\tau^{-\theta-\frac{\theta}{\gamma}}\,, \label{38} \ee
where $\text{B}(x,y)$ is the Beta function. For any fixed
$1<\gamma<2$ and $0<\theta<1$, there is a
$\tau^*(\gamma,\theta)<\infty$ such that \be \left|\delta
\frac{d^\theta \mathbb{R}}{d\tau^\theta} \right|\ll
\beta\,\mathbb{R}^\gamma\simeq\tau^{-\theta} \qquad
\text{if}\qquad \tau\gg \tau^*(\gamma,\theta) \ee so that
the quasistatic approximation becomes applicable. The
physical background of the power asymptotics (\ref{33}) of
the solution of the linear equation (\ref{27}) and of the
quasistatic approximation (\ref{36}) of the nonlinear
equation (\ref{19}) is obvious: the asymptotics
$\mathbb{R}\sim \tau^{-\theta}$ given by (\ref{33}) is a
consequence of the power tail $\Phi(t)\sim t^{-\theta-1}$
of the bare Omori law, while the more slowly decaying
$\mathbb{R}\sim \tau^{-\theta/\gamma}$ given by (\ref{36})
is the result of an interplay between the long memory
property of the bare Omori law and the amplification by
the power law $\Omega(z)\sim z^\gamma$, a signature of the
broad distribution of productivities of daughter
aftershocks from mother earthquakes. This gives rise to a
renormalization of the exponent $\theta$ into  a smaller
exponent $\theta/\gamma$ (for $1<\gamma<2$).

\subsection{PDF of the total duration of aftershocks branching
processes}

The previous sections have discussed in details how to obtain the
complementary CDF $\mathbb{R}$ of the total duration of
aftershock branching processes, corresponding to some
first generation aftershock, triggered by a main
earthquake. The CDF $\mathbb{P}_\mu$ of the total duration of
aftershock triggering processes, taking into account all
aftershocks triggered by a main earthquake of fixed magnitude, is described
by relation (\ref{11}). The corresponding PDF of the total duration
of an aftershock sequence is thus equal to
\be
\mathbb{W}_\mu(\tau)=-\mu\kappa
e^{-\kappa\mu\mathbb{R}(\tau)}
\frac{d\mathbb{R}(\tau)}{d\tau}\,.
\label{48}
\ee
If $\mu\kappa\gg 1$ (as is the case for a large earthquake which has
a large average productivity), then, due to the exponential factor in
(\ref{48}), this PDF differs significantly from zero only if
$\mathbb{R}$ is very small. Then using the expression for small values of
$\mathbb{R}$ described by the quasistatic
approximation (\ref{36}), we obtain
\be
\mathbb{W}_\mu(\tau)=\frac{d\mathbb{P}_\mu(\tau)}{d\tau}
\simeq \frac{\theta\mu\kappa}{\gamma\beta^{1/\gamma}}
~\tau^{-1-\theta/\gamma}~ \exp\left(
-\frac{\mu\kappa}{\beta^{1/\gamma}}\,
\tau^{-\theta/\gamma}\right)\,.
\label{49}
\ee
Expression (\ref{49}) is our main result.
Fig.~4 shows a log-log plot of the
PDF (\ref{49}) for different values of the mainshock size $\mu\kappa$ for
$\gamma=1.25$ and $\theta=0.2$
(Recall that $\beta$ is given by (\ref{betadef}) and we put it equal to $1$
to draw Fig.~4).

Expression (\ref{49}) shows that the power law tail holds for
durations $t/c > t_{\mu}/c \propto (\mu \kappa)^{\gamma / \theta}
\sim 10^{(\alpha \gamma/\theta) m}$ for
which the exponential factor goes to $1$. Thus, for
$\theta$ small ($\approx 0.1-0.3$ as seem to be relevant for earthquakes),
expression (\ref{49}) exhibits a very strong dependence on the mainshock
magnitude through its impact (\ref{mudef}) on the mark $\mu$.
Therefore, the most
relevant part of
the distribution of the durations for small mainshocks is controlled by
the power law tail $\tau^{-1-\theta/\gamma}$. In contrast, the observable
part of the distribution of durations for very large
mainshocks is controlled by the exponential term which, together with
the power law prefactor, leads to a maximum: for very large $\mu$,
$\mathbb{W}_\mu(\tau)$
starts from zero for $\tau=0$ and then
increase up to a maximum before crossing over slowly to the power law tail
$\tau^{-1-\theta/\gamma}$, as illustrated in Fig.~4.

\subsection{Crossover from critical to subcritical regime}

The asymptotics of the complementary CDF
$\mathbb{R}$ satisfies the equation (\ref{35}) in the quasistatic
approximation. In the
subcritical regime, using the polynomial approximation (\ref{15}),
one can rewrite equation (\ref{35}) in the form
\be
(1-n)\mathbb{R}+\beta \mathbb{R}^\gamma=\tau^{-\theta}\,.
\label{44}
\ee
It is seen from this equality that if
$\mathbb{R}>\mathbb{R}_c$, where
\be
\mathbb{R}_c=\left(\frac{1-n}{\beta}\right)^g ~,
\label{45}
\ee
then one can neglect the linear term in the left-hand-side of equality
(\ref{44}) and obtain the power law (\ref{36}), typical of the critical
regime $n=1$. In contrast, if
$\mathbb{R}<\mathbb{R}_c$, then the subcritical scenario of
the complementary CDF $\mathbb{R}$ dominates and
equality (\ref{44}) gives the subcritical power law
\be
\mathbb{R}\simeq \frac{\tau^{-\theta}}{1-n}\,.
\label{46}
\ee
It follows from (\ref{44}) and (\ref{45}) that the time of the crossover from
the critical to the subcritical regime is equal to
\be
\tau_c\simeq \left(
\frac{\beta^g}{(1-n)^{g+1}}\right)^{1/\theta}\,.
\label{47}
\ee

{}

\clearpage

\begin{quote}
\centerline{
\includegraphics[width=10cm]{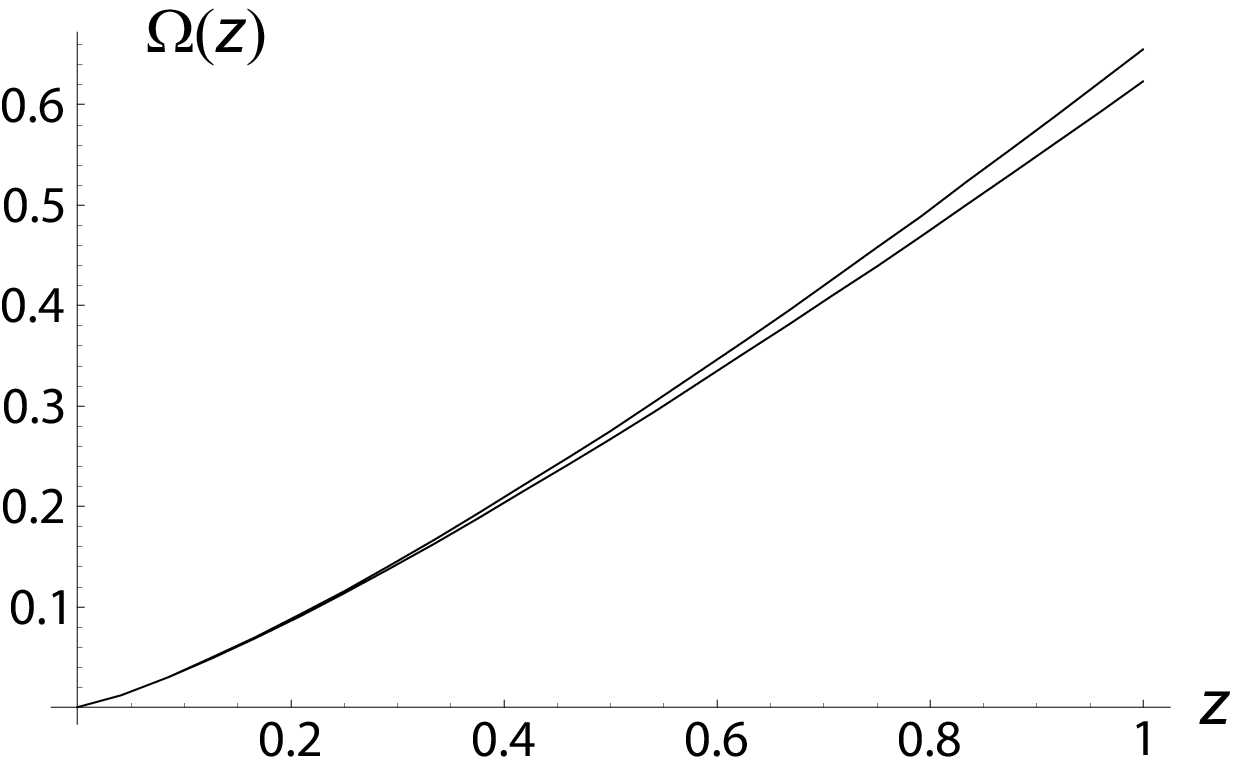}
}
{\bf Fig.~1:} \small{Plots of exact $\Omega(z)$ defined by
(\ref{14}) (lower curve) and its
pure power approximation (\ref{16}) (upper curve) for $\gamma=1.25$ and $n=1$.}
\end{quote}

\clearpage

\begin{quote}
\centerline{\includegraphics[width=10cm]{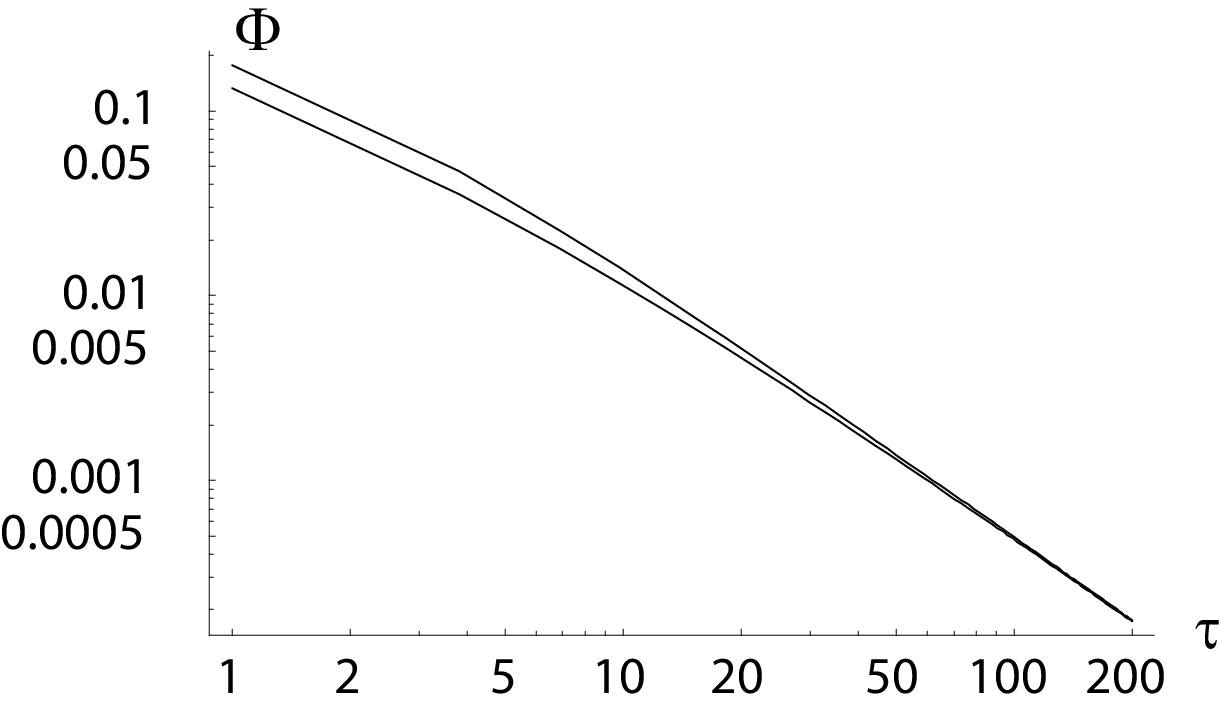} }
{\bf
Fig.~2:} \small{Loglog plots of the direct Omori law
$\Phi(t)$ defined
in (\ref{eq1dffrf}) (lower curve) and of the fractional exponential
distribution
(\ref{20}) (upper curve) for
$\theta=0.5$ and $c=1$. }
\end{quote}

\clearpage

\begin{quote}
\centerline{
\includegraphics[width=10cm]{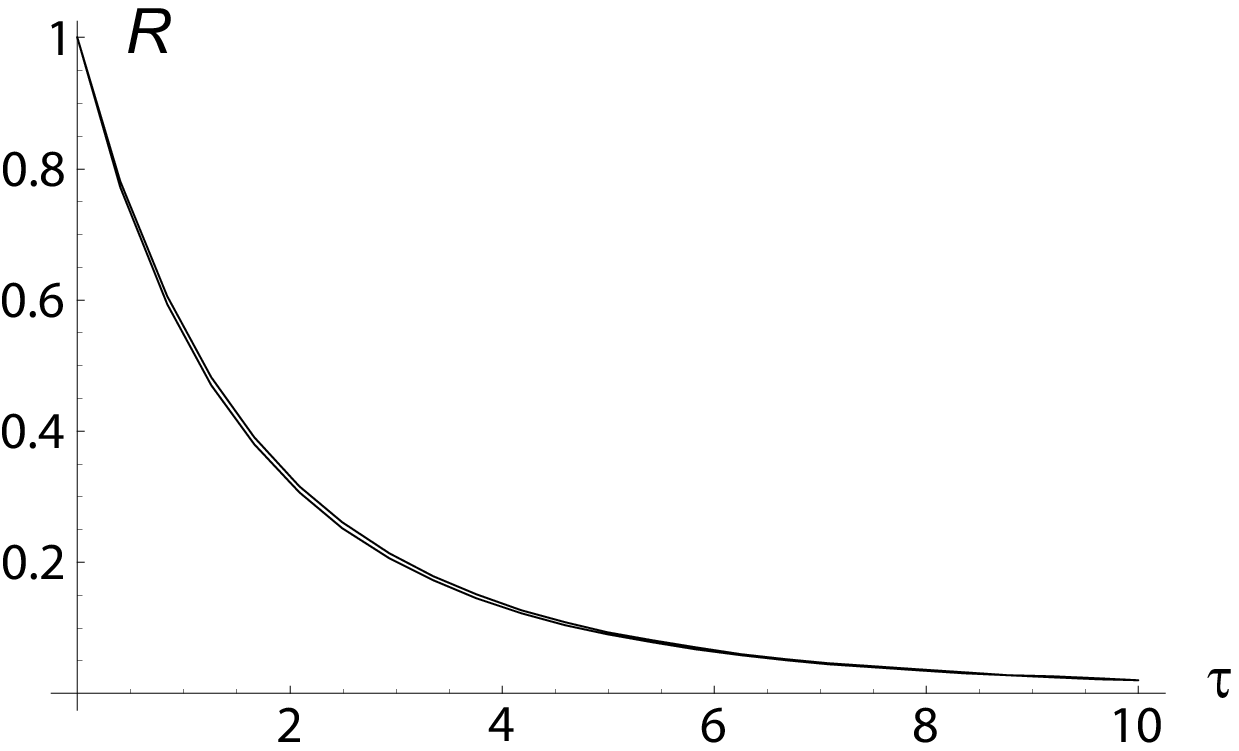}
}
{\bf Fig.~3:} \small{Plot of the numerical solution of
equation (\ref{24}) for the complementary CDF $\mathbb{R}$ of the
total duration of an aftershock sequence and the corresponding
analytical approximate expression (\ref{25}) for $\mathbb{R}$ for
the parameters $\gamma=1.25$ and $n=0.99$.}
\end{quote}

\clearpage

\begin{quote} \centerline{
\includegraphics[width=10cm]{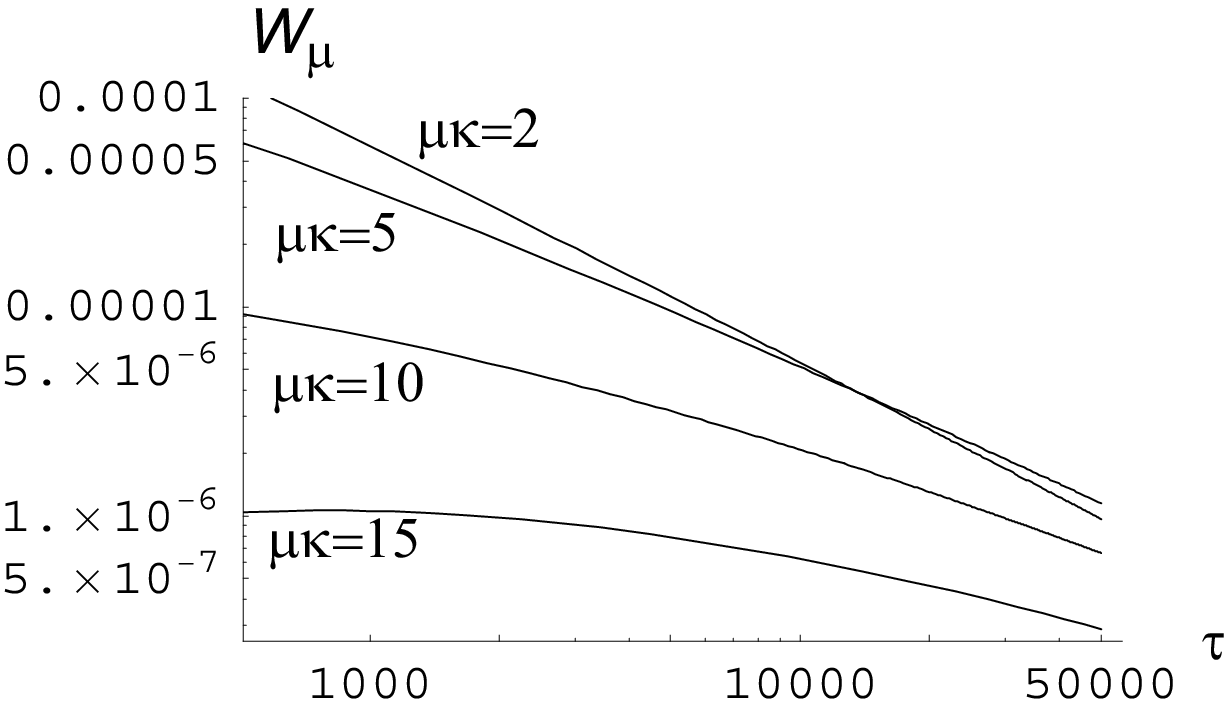}
}
{\bf Fig.~4:} \small{Log-log plots of the PDF
(\ref{49}) of the total aftershock sequence durations for a mainshock
of mark $\mu$, with $\mu\kappa=2, 5, 10, 15$, for the parameter values
$\gamma=1.25$, $\theta=0.2$ and $n=1$.}
\end{quote}

\end{document}